\documentclass[letterpaper,11pt]{article}
\pdfoutput=1

\usepackage{jheppub}
\usepackage{booktabs} 
\usepackage{graphicx} 
\usepackage{siunitx}
\usepackage{mathrsfs}
\usepackage[T1]{fontenc}
\usepackage{cleveref}

\usepackage{bm}
\usepackage{bbm}
\usepackage{xspace}

\usepackage{amsmath,latexsym}
\usepackage{pstricks}
\usepackage{color}

\begin{document}
	
	
	\title{An effective field theory of damped ferromagnetic systems}
	
	\author{Jingping Li}
	\affiliation{Department of Physics, Carnegie Mellon University, Pittsburgh, PA 15213}
	
	\emailAdd{jingpinl@andrew.cmu.edu}

	\vspace{0.3cm}


\abstract{
Using the in-in formalism, we generalize the recently constructed
magnetoelastic EFT \cite{PavaskarPencoRothstein} to describe the
damping dynamics of ferromagnetic systems at long wavelengths. We
find that the standard Gilbert damping term naturally arises as the
simplest leading-order symmetry-consistent non-conservative contribution
within the in-in framework. The EFT is easily generalized to scenarios
with anisotropy and inhomogeneity. In particular, we find the classic
Landau-Lifshitz damping term emerges when isotropy is broken by a
constant external background field. This provides a first principle
explanation for distinguishing the two types of damping dynamics that
were originally constructed phenomenologically. Furthermore, the EFT
framework could also incorporate intrinsic anisotropy of the material
in a straightforward way using the spurion method. For systems with
inhomogeneity such as nontrivial spin textures, we find that the leading
order derivative correction yields the generalized Gilbert damping
equations that were found in condensed matter literature. This shows
that the EFT approach enables us to derive the form of higher-derivative-order
corrections in a systematic way. Lastly, using the phonon-magnon
coupling deduced in the magnetoelastic EFT, we are able to make a
prediction for the generic form of the phononic contribution to the
damping equation.}
\maketitle	

\section{Introduction}

It has long been established that the methodology of coset construction
serves as a powerful tool of relativistic effective field theories
(EFTs) of Goldstone bosons (e.g. pions from spontaneously broken approximate
chiral symmetry) \cite{ColemanWessZumino,Callanetal,Volkov,Ogievetsky}.
In recent years, many works have demonstrated that its versatility
is extendable to condensed matter systems where we are interested
in the macroscopic behavior which are usually the massless low energy
excitations \cite{EFT,CMEFT}. Using this approach, a recent paper
\cite{PavaskarPencoRothstein} constructed an EFT of magnetoelastic
systems where the phonons and magnons are considered Goldstones associated
with translations spontaneously broken by the ground state location
of the material (the lattice) and an $SO(3)$ symmetry of the magnetic
moments by their ground state orientations. The EFT approach provides a systematic way to understand phonon-magnon interactions from first principles and predict the forms of higher-order corrections which has not been done previously.

While the paper was focused on conservative dynamics, there has also
been extensive study on the theoretical description of non-conservative
dynamics of damped magnetic systems since the seminal work of Landau,
Lifshitz, and later Gilbert \cite{LandauLifshitz,Gilbert}. However,
to our knowledge, the prior works were mostly model-dependent phenomenological
descriptions. It is therefore desirable to have a first principle
derivation from a similar many-body EFT perspective.

On the other hand, the Schwinger-Keldysh formalism \cite{Schwinger,Keldysh}
(and the related in-in formalism \cite{Galley,GalleyTsangStein})
has been known to describe the quantum field theory of open systems
and hence fully capable of describing dissipative effects. In recent
years, its power has been successfully extended to the EFT framework
of dissipative dynamics in astrophysics and black holes \cite{EndlichPenco,GoldbergerRothstein,GoldbergerLiRothstein}
that systematically derives dissipative equations of motion. Therefore,
it is natural to consider its utility in describing deriving damping
equations in condensed matter EFTs.

In this paper, combining the power of the two techniques, we apply
the Schwinger-Keldysh formalism to incorporate dissipative effects
into the EFT of magnons to reproduce the known results of magnetic
damping. In section 2, we review the coset construction of the magnon
EFT as well as the techniques in Schwinger-Keldysh formalism to be
applied in this paper. Section 3 derives the original Gilbert damping
equation for homogeneous and isotropic materials. In section 4, we
move on to more general materials and recover the Landau-Lifshitz
damping equation for anisotropic systems and generalized Gilbert damping
for spatially inhomogeneous materials. In section 5, we derive the
damping terms originating from the magnon-phonon interaction.

Conventions: we use natural units where $\hbar=1$. Unless specified,
the uppercase Latin indices $A,B,C\dots$ denote the full internal
spin symmetry space which runs over $1,2,3$ while the lower case
ones $a,b,c\dots$ in the begining alphabet index the broken subspace
$1,2$. Those in the middle alphabet $i,j,k\dots$ run over the three
spatial dimensions (to generalize to higher dimensions, the internal
spin symmetry will have to be modified accordingly).

\section{Review of the magnon EFT and Schwinger-Keldysh formalism}

In this section, we first provide a self-contained review on the symmetries
and the corresponding coset construction of magnon-phonon EFT proposed
by \cite{PavaskarPencoRothstein}. Furthermore, we summarize the Schwinger-Keldysh
formalism which is the central tool for deriving the dissipative equations
of motion.

\subsection{Symmetry breaking in magnetoelastic systems}

We follow the derivations in {[}1{]}. The symmetries under consideration
are the spatial Galilean group (generated by translations $P_{i}$,
rotations, $L_{i}$, boost $K_{i}$) and internal symmetries (internal
translations $T_{i}$, internal rotations $Q_{i}$, spin rotations
$S_{A}$). The algebra of the generators is given by
\begin{align}
\left[L_{i},K_{j}\right]=i\epsilon_{ijk}K_{k}, & \left[L_{i},P_{j}\right]=i\epsilon_{ijk}P_{k},\\
\left[K_{i},H\right]=-iP_{k}, & \left[K_{i},P_{j}\right]=-iM\delta_{ij},\\
\left[Q_{i},T_{j}\right]=i\epsilon_{ijk}T_{k}, & \left[Q_{i},Q_{j}\right]=i\epsilon_{ijk}Q_{k},\\
\left[S_{A},S_{B}\right]=i\epsilon_{ijk}K_{k}, & \left[L_{i},L_{j}\right]=i\epsilon_{ijk}L_{k}.
\end{align}
In particular, the $T_{i}$ and $Q_{i}$ generators generate translations
and rotations on the ``comoving'' coordinates $\phi^{I}(x)$ (or
the Lagrangian coordinates in continuum mechanics which simply gives
an initial labeling to the continuum)
\begin{equation}
\phi^{I}(x)\rightarrow\phi^{I}(x)+a^{I},\ \phi^{I}(x)\rightarrow R_{\ J}^{I}\phi^{J}(x),
\end{equation}
and $S_{A}$ generates the internal rotation on the orientation of
the spin 
\begin{equation}
\mathcal{N}^{A}\rightarrow O_{\ B}^{A}\mathcal{N}^{B},
\end{equation}
where $O=e^{i\chi_{a}S^{a}}$ and the N\'eel vector $\vec{\mathcal{N}}$
is the order parameter for the spin orientation.

In the ground state, the order parameters gain vacuum expectation
values (VEVs) which we choose to be
\begin{equation}
\left\langle \vec{\phi}(x)\right\rangle =\vec{x},\ \left\langle \vec{\mathcal{N}}\right\rangle =\hat{x}_{3}.
\end{equation}
The VEVs are not invariant under the transformations and hence spontaneously
break some of the symmetries
\begin{equation}
Unbroken=\begin{cases}
H\\
P_{i}+T_{i}\equiv\bar{P}_{i}\\
L_{i}+Q_{i}\\
S_{3}\\
M
\end{cases},\ Broken=\begin{cases}
K_{i}\\
T_{i}\\
Q_{i}\\
S_{1},S_{2}\equiv S_{a}
\end{cases}.
\end{equation}
The parametrization of the ground state manifold which is simply the
broken symmetry transformations plus the unbroken translations is
given by
\begin{equation}
\Omega=e^{-itH}e^{ix^{i}\bar{P}_{i}}e^{i\eta^{i}K_{i}}e^{i\pi^{i}T_{i}}e^{i\theta^{i}Q_{i}}e^{i\chi^{a}S_{a}},
\end{equation}
where $\eta^{i}$, $\theta^{i}$, $\chi^{a}$, and $\pi^{i}=\phi^{i}-x^{i}$
are the corresponding Goldstone fields.

For any Goldstone fields $\psi^i$ corresponding to the broken generators $X_i$, their covariant derivatives are the basic building blocks of the low energy EFT. They are systematically computed by the coset construction using the Mauer-Cartan forms of the broken group
\begin{equation}
\Omega^{-1}\partial_{\mu}\Omega\supset\left(\nabla_{\mu}\psi^{i}\right)X_{i},
\end{equation}
by computing the coefficients of the broken generator $X_{i}$. In
addition, in the case that one broken generator $X'$ appears in the
commutation algebra of the other $X$' with the unbroken translations
\begin{equation}
\left[\bar{P},X\right]\supset X',
\end{equation}
it means that the two Goldstones are not independent and one of them
can be eliminated. This is known as the inverse Higgs phenomenon.

The result of this exercise is that $\eta^{i}$, $\theta^{i}$ are
eliminated and the only independent degrees of freedom are the magnons
$\chi^{a}$ and phonons $\pi^{i}$, and at leading order in derivatives,
they appear in the following combinations:
\begin{align}
\nabla_{(i}\pi_{j)} & =(D\sqrt{D^{T}D}D^{-1})_{ij}-\delta_{ij},\\
\nabla_{t}\chi^{a} & =\frac{1}{2}\epsilon^{aBC}\left\{ O^{-1}\left[\partial_{t}-\partial_{t}\pi^{k}(D^{-1})_{k}^{\ j}\partial_{j}\right]O\right\} _{BC},\\
\nabla_{i}\chi^{a} & =\frac{1}{2}\epsilon^{aBC}\left(O^{-1}\partial_{i}O\right)_{BC},
\end{align}
where $D_{ij}=\delta_{ij}+\partial_{i}\pi_{j}$. \cite{PavaskarPencoRothstein}
found the most general action for ferromagnetic material in the form
\begin{align}
\mathcal{L}=&\frac{c_{1}}{2}\det\left(D\right)\epsilon^{ab}\left[\left(O^{-1}\partial_{t}O\right)_{ab}-\partial_{t}\pi^{k}\left(D^{-1}\right)_{k}^{\ j}\left(O^{-1}\partial_{j}O\right)_{ab}\right]\\&-\frac{1}{2}F_{2}^{ij}(\nabla_{(i}\pi_{j)})\nabla_{i}\chi_{a}\nabla_{j}\chi^{a}-\frac{1}{2}F_{3}(\nabla_{(i}\pi_{j)})\nabla_{t}\chi_{a}\nabla_{t}\chi^{a},\label{eq:mag}
\end{align}
where the first term is similar to a Wess-Zumino-Witten (WZW) term
that differs by a total derivative under the symmetry transformation.

\subsection{The magnon EFT and the conservative equation of motion}

To derive the equations of motion for magnons, it is more convenient
to express the magnon fields in the nonlinear form
\begin{equation}
\hat{n}=\mathcal{R}(\chi)\hat{x}_{3}=\left(\sin\theta\cos\phi,\sin\theta\sin\phi,\cos\theta\right),
\end{equation}
where the two magnon fields are related to the angular fields by 
\begin{equation}
\chi_{1}=\theta\sin\phi,\quad\chi_{2}=\theta\cos\phi.
\end{equation}
Physically, this unit vector represents the direction of the magnetic
moment. Under this representation, the pure magnon Lagrangian (in
the absence of phonon excitations) becomes
\begin{equation}
\mathcal{L}\rightarrow \frac{c_2}{2}\epsilon^{ab}\left(O^{-1}\partial_t O\right)_{ab}+\frac{c_{6}}{2}\left(\partial_{t}\hat{n}\right)^{2}-\frac{c_{7}}{2}\left(\partial_{i}\hat{n}\right)^{2},\label{eq:magnon}
\end{equation}
where $F_{3}(0)=c_{6}$ and $F_{2}^{ij}(0)=c_{7}\delta^{ij}$.

The dispersion relation for the quadratic Lagrangian has two solutions \cite{PavaskarPencoRothstein}
\begin{equation}
\omega_+^2=\bigg(\frac{c_2}{c_6}\bigg)^2+O(k^2),\quad \omega_-^2=\bigg(\frac{c_7}{c_2}\bigg)^2k^4+O(k^6).\label{eq:disp}
\end{equation}
For ferromagnetic materials, where $c_2~(c_6c_7)^{3/4}$, the first mode is gapped around the EFT cutoff scale, while the second has $\omega\sim k^2$ scaling and exits the EFT. In the long wavelength limit, we may assign the scaling $\partial_t^{1/2} \sim \partial_i$ to the derivatives for ferromagnets.

To derive the equation of motion, we notice that an action of this form has a symmetry under the infinitesimal spin
rotation $\delta\hat{n}=\vec{\omega}\times\hat{n}$, where $\vec{\omega}$
is the constant infinitesimal parameter. It can be shown that the
Wess-Zumino-Witten term contributes to a total derivative $\partial_{\mu}\vec{F}^{\mu}$
under this transformation. Using the equation for Noether current
in Lagrangian mechanics 
\begin{equation}
\vec{J}^{\mu}=\hat{n}\times\frac{\partial\mathcal{L}}{\partial\partial_{\mu}\hat{n}}-\vec{F}^{\mu},\label{eq:curr}
\end{equation}
we find the conserved current
\begin{equation}
\vec{J}^{0}=-c_{2}\hat{n}-c_{6}\partial_{t}\hat{n}\times\hat{n},\ \vec{J}^{i}=c_7\partial_{i}n\times\hat{n}.\label{eq:curr-1}
\end{equation}
The continuity equation $\partial_{\mu}\vec{J}^{\mu}=0$ is then explicitly
\begin{equation}
c_{2}\partial_{t}\hat{n}=-\left(c_{6}\partial_{t}^{2}\hat{n}-c_{7}\nabla^{2}\hat{n}\right)\times\hat{n},\label{eq:cons}
\end{equation}
which is the equation of motion for Landau-Lifshitz model of magnetism.

\subsection{The Schwinger-Keldysh formalism\label{subsec:The-Schwinger-Keldysh-formalism}}

The appropriate formalism for non-conservative system is the so-called
in-in or Schwinger-Keldysh formalism \cite{Schwinger,Keldysh,Galley,GalleyTsangStein}.
The basic idea is that there is an external sector $X$ that the energy
is dissipated into since the total energy needs to be conserved. The
external sector could evolve into any final state which we do not
observe, so all the dynamics are inclusive of the final states in
the Hilbert space of $X$
\begin{equation}
\sum_{X_{out}}\langle X_{in}\vert\dots\vert X_{out}\rangle\langle X_{out}\vert\dots\vert X_{in}\rangle\equiv\langle\dots\rangle_{in},
\end{equation}
and depends only on the initial state (hence the name in-in). We can
generate an effective action for the in-in observables via the Schwinger-Keldysh
closed time path integral

\begin{equation}
\exp\bigg[i\Gamma[q;\tilde{q}]\bigg]=\int_{initial}\mathcal{D}X\mathcal{D}\tilde{X}\exp\bigg[iS[q,X]-iS[\tilde{q},\tilde{X}]\bigg],\label{eq:skp}
\end{equation}
where we are integrating over an additional copy of variables $\tilde{X}$
which corresponds to evolving back to the boundary conditions fixed
at the initial time.

The equation of motion for the degrees of freedom in the observed sector $q$ can be derived from the action functional
$\Gamma[q;\tilde{q}]$  by
\begin{equation}
\frac{\delta}{\delta q}\Gamma[q;\tilde{q}]\Biggr\vert_{q=\tilde{q}}=0.
\end{equation}
Any external sector operator $\mathcal{O}(X)$ coupled to some operator in the observable
sector $F(q)$ by the interaction term $\int dx\mathcal{O}\left(X(x)\right)F\left(q\left(x'\right)\right)$
(where $x$ is the corresponding spacetime coordinates) would enter
the equations of motion in terms of 
\begin{equation}
\langle\mathcal{O}\left(X(x)\right)\rangle_{in}=\int_{initial}\mathcal{D}X\mathcal{D}\tilde{X}\exp\bigg[iS[q,X]-iS[\tilde{q},\tilde{X}]\bigg]\mathcal{O}(X),\label{eq:skp-1}
\end{equation}
where we have abbreviated $\langle\mathcal{O}(X)\rangle_{in}\equiv\langle X_{in}\vert\mathcal{O}(X)\vert X_{in}\rangle$.
Just as in the perturbative quantum field theory correlation functions
calculated by Feynman propagators, this can be similarly calculated
using the Schwinger-Keldysh propagators
\begin{equation}
\langle\mathcal{O}_{a}(x)\mathcal{O}_{b}(x')\rangle=\left(\begin{array}{cc}
\langle T\mathcal{O}(x)\mathcal{O}(x')\rangle & \langle\mathcal{O}(x')\mathcal{O}(x)\rangle\\
\langle\mathcal{O}(x)\mathcal{O}(x')\rangle & \langle\tilde{T}\mathcal{O}(x)\mathcal{O}(x')\rangle
\end{array}\right),
\end{equation}
where $T$ and $\tilde{T}$ represent time and anti-time orderings.
The sub-indices label the first and the second copy, which determines
the relative time-ordering of the operators.

Explicitly, the linear response gives
\begin{equation}
\langle\mathcal{O}(x)\rangle=i\int dx'\{\langle T\mathcal{O}(x)\mathcal{O}(x')\rangle-\langle\mathcal{O}(x')\mathcal{O}(x)\rangle\}F\left(q(x')\right)+\mathcal{O}(F^{2}).
\end{equation}
or equivalently
\begin{equation}
\langle\mathcal{O}(x)\rangle=\int dx'G_{R}(x,x')F\left(x'\right),
\end{equation}
with the retarded Green's function given by
\begin{align}
G_{R}(x,x') & =i\theta(t-t')\langle[\mathcal{O}(x),\mathcal{O}(x')]\rangle\nonumber \\
 & =i(\langle T\mathcal{O}(x)\mathcal{O}(x')\rangle-\langle\mathcal{O}(x')\mathcal{O}(x)\rangle).
\end{align}
Therefore, the exact form of the damping term in the equation of motion would depend on the detailed structure of these retarded response functions.

\section{Gilbert damping term from EFT}

\subsection{Coupling at the leading order}

The composite operators $\mathcal{O}^{r}(X)$ that encapsulate the
external sector transform under arbitrary representations (labeled
by $r$), provided they form invariants of the unbroken $SO(2)$ with
the magnon $\chi_{a}$ and derivatives. In order to achieve this,
the operators have to be dressed with the broken $SO(3)/SO(2)$ subgroup
parametrized by the Goldstones $T^{R}(\chi)$ in the corresponding
representation
\begin{equation}
\tilde{\mathcal{O}}^{r}(X)\equiv\mathcal{R}^{r}(\chi)\mathcal{O}^{r}(X),
\end{equation}
such that they transform covariantly under the unbroken subgroup \cite{Delacretazetal}.

In the long wavelength regime, the theory is organized by a spatial derivative expansion. In fact, the simplest invariant operator at zeroth-order in the derivative expansion is the singlet aligned along the ground state orientation $\hat{x}_{3}$ 
\begin{equation}
S_{int}=\int d^{4}x\tilde{\mathcal{O}}^{3}(X)\equiv\int d^{4}x\hat{x}_{3}\cdot\tilde{\vec{\mathcal{O}}}(X),
\end{equation}
(note that we are adopting manifestly relativistic notations for spacetime
and energy-momentum for convenience, albeit the system may or may
not be relativistic). Equivalently, we may write
\begin{equation}
S_{int}=\int d^{4}x\hat{n}\cdot\vec{\mathcal{O}}(X)\label{eq:diss}
\end{equation}
where $\hat{n}=O(\chi)\hat{x}_{3}$ as defined previously.

 At the same order in this expansion, there could
be more operators that can be added, such as when the operator is
a two-index tensor operator $\mathcal{O}^{(2)}$ and we may have combinations
of the form $\hat{n}\cdot\mathcal{O}^{(2)}\cdot\hat{n}$. However,
as will be explained in the next subsection, due to the constraint
$\hat{n}^{2}=1$, these will not lead to any new contributions, and
from the perspective of the EFT, they are redundant operators. Hence,
Eq. (\ref{eq:diss}) is the only non-trivial operator at this order.

Furthermore, we observe that the form of the operator restricts the
possible external sector that can couple in this way. For example,
one can form such operators from fermions, where
\begin{equation}
\vec{\mathcal{O}}(\psi)=\psi^{\dagger}\vec{\sigma}\psi+\psi\vec{\sigma}\psi^{\dagger},\label{eq:ferm}
\end{equation}
where $\vec{\sigma}$ are the Pauli matrices acting as the intertwiner
between the spinor and $SO(3)$ spin space. On the other hand, phonons
cannot form operators in this form, and hence will not contribute
in this way.

\subsection{Gilbert damping}

The undamped equation of motion is derived from the continuity equation
\begin{equation}
\partial_{\mu}\vec{J}^{\mu}=0
\end{equation}
corresponding to the spin rotation transformation $\delta\hat{n}=\vec{\omega}\times\hat{n}$.
To derive the non-conservative equation of motion, we need to find
out how the additional term Eq. (\ref{eq:diss}) affects the continuity
equation.

The spin rotation transformation on $\hat{n}$ alone is itself a valid
symmetry for the pure magnon action Eq. (\ref{eq:magnon}), but the
interaction term Eq. (\ref{eq:diss}) is not invariant if we keep
the external sector fixed. A standard trick of Noether theorem is
that for an arbitrary symmetry transformation $\phi(x)\mapsto\phi(x)+f(x)\epsilon$,
if we promote the global symmetry variation parameter to be an arbitrary
local variation $\epsilon(x)$, the total variation takes the form
\begin{equation}
\delta S=-\int d^{4}xJ^{\mu}\partial_{\mu}\epsilon,
\end{equation}
such that when $\epsilon$ is a constant, invariance under the symmetry
transformation is guaranteed $\delta S=0$ even off-shell (equation
of motion is not satisfied). Integrate by parts, we find that on-shell
\begin{equation}
\delta S=\int d^{4}x\left(\partial_{\mu}J^{\mu}\right)\epsilon.
\end{equation}
However, for an arbitrary $\epsilon(x)$, this is also the equation
of motion, since
\begin{equation}
\delta S=\int d^{4}x\frac{\delta S}{\delta\phi}f(x)\epsilon(x).
\end{equation}
Therefore, the effect of an additional term in the action $\Delta S$
is adding a term to the current divergence
\begin{equation}
\partial_{\mu}J^{\mu}=0\rightarrow\partial_{\mu}J^{\mu}+\frac{\delta\Delta S}{\delta\phi}f(x)=0.
\end{equation}

For the spin rotations, the variation of the pure magnon EFT with
local $\omega(x)$ is given by
\begin{equation}
\delta S=\int d^{4}x\left(\partial_{\mu}\vec{J}^{\mu}\right)\cdot\vec{\omega}.
\end{equation}
Correspondingly, the addition of Eq. (\ref{eq:diss}) leads to the
modification

\begin{equation}
\delta S=\int d^{4}x\left(\partial_{\mu}\vec{J}^{\mu}+\hat{n}\times\vec{\mathcal{O}}(X)\right)\cdot\vec{\omega}.
\end{equation}
Thus, the (non-)continuity equation becomes
\begin{equation}
\partial_{\mu}\vec{J}^{\mu}=-\hat{n}\times\vec{\mathcal{O}}(X).
\end{equation}
When we focus on the measurements of the magnons, the effect of the
external sector enters as an in-in expectation value $\left\langle \vec{\mathcal{O}}\right\rangle _{in}$.
This may be evaluated via the in-in formalism, and the leading order
contribution is given by
\begin{equation}
\vec{\omega}\cdot\int d^{4}x\partial_{\mu}\vec{J}^{\mu}=-\vec{\omega}\cdot\int d^{4}x\hat{n}\times\left\langle \vec{\mathcal{O}}\right\rangle _{in}=-\vec{\omega}\cdot\int d^{4}x\hat{n}\times\int d^{4}x'\left(G_{R}\cdot\hat{n}'\right),
\end{equation}
where $G_{R}(t',\vec{x}';t,\vec{x})$ is the retarded response function
of the operator $\vec{\mathcal{O}}$.

In frequency space, we have $G_{R}(t,\vec{x})=\int\frac{d^{3}\vec{k}d\omega}{(2\pi)^{4}}e^{-i\omega t+\vec{k}\cdot\vec{x}}\tilde{G}_{R}(\omega,\vec{k})$
and furthermore, using the spectral representation (making the spin
space indices explicit temporarily)
\begin{equation}
\tilde{G}_{R}^{AB}(\omega,\vec{k})=\int_{-\infty}^{\infty}\frac{d\omega_{0}}{\pi}\frac{i}{\omega-\omega_{0}+i\epsilon}\rho^{AB}(\omega_{0},\vec{k}),
\end{equation}
we can separate the prefactor using the identity
\begin{equation}
\frac{i}{\omega-\omega_{0}+i\epsilon}=\pi\delta(\omega-\omega_{0})+\mathcal{P}\frac{i}{\omega-\omega_{0}},
\end{equation}
 into a $\delta$-function and a principal part. The dissipative part
is captured by the former
\begin{equation}
\tilde{G}_{R,diss}^{AB}(\omega,\vec{k})=\int_{-\infty}^{\infty}d\omega_{0}\delta(\omega-\omega_{0})\rho^{AB}(\omega_{0},\vec{k})=\rho^{AB}(\omega,\vec{k}).
\end{equation}
The indices of this spectral function lives in the spin $SO(3)$ space
and, for isotropic systems, should be built from invariant tensors
$\delta^{AB}$, $\epsilon^{ABC}$. However, the latter could not neither
form a two-index object nor respect parity invariance by itself, so
the only symmetry-consistent possibility is
\begin{equation}
\rho^{AB}(\omega,\vec{k})=f(\omega,\vert\vec{k}\vert^{2})\delta^{AB},\label{eq:spec}
\end{equation}
with $f$ assumed to be an analytic function of its arguments such
that it has a smooth limit as $\omega$ goes to zero. Dissipative
dynamics is antisymmetric under time reversal, so it should be odd
under the simultaneous transformation $\omega\leftrightarrow-\omega$
and $(A,B)\leftrightarrow(B,A)$, meaning that the leading order contribution
is given by 
\begin{equation}
\rho^{AB}(\omega,\vec{k})=-i\mathscr{C}\omega\delta^{AB}\label{eq:lospec}
\end{equation}
or in real spacetime
\begin{equation}
G_{R,diss}^{AB}(t,\vec{x})=\mathscr{C}\frac{\partial}{\partial t}\delta(t)\delta^{3}(\vec{x})\delta^{AB}.\label{eq:resp}
\end{equation}
$\mathscr{C}$ could be understood as a Wilson coefficient in this
non-conservative sector.

From this, we arrive at the equation
\begin{equation}
\partial_{\mu}\vec{J}^{\mu}=-\mathscr{C}\hat{n}\times\frac{\partial}{\partial t}\hat{n}.
\end{equation}
Combining with the conservative part of the continuity equation Eq.
(\ref{eq:cons}), we find the Gilbert damping equation
\begin{equation}
\frac{\partial}{\partial t}\vec{m}=-\gamma\vec{m}\times\frac{\partial}{\partial t}\vec{m}+\dots,\label{eq:gilb}
\end{equation}
where $\gamma=\mathscr{C}/(c_{2}m_{s})$, $m_{s}\hat{n}=\vec{m}$
(for uniform materials), and the higher-order terms on the right-hand
side of the original equation of motion Eq. (\ref{eq:cons}) are contained
in$\dots$ which will be omitted in the following.

When we have another singlet of the form $\hat{n}\cdot\mathcal{O}^{(2)}\cdot\hat{n}$,
the effect of the extra $\hat{n}$s is a replacement of the spectral
density $\rho^{AB}\rightarrow\rho^{ACBD}\hat{n}_{C}\hat{n}_{D}$.
The tensor basis still consists of Kronecker delta since the only
structure that Levi-Civita tensors could contract to $\hat{n}$ and
would vanish automatically. Therefore, any additional structures will
appear in the form of the inner product $\hat{n}\cdot\hat{n}$ due
to contractions with the Kronecker delta. Consequently, they do not
lead to anything new due to the normalization condition $\hat{n}^{2}=1$.

We see that using the in-in formalism, the form of the damping equation
and the coefficients are completely fixed by the principles of EFT:
the symmetries, power counting, and Wilson coefficients.

\section{More general materials}

By choosing the spectral function to depend only on invariant tensors
and the frequency, we naturally arrive at the Gilbert damping Eq.
(\ref{eq:gilb}) which applies to isotropic and homogeneous systems.
However, in generic materials, we may be interested in situations
with more general materials which for instance have non-trivial spin
textures or highly anisotropic lattices and thus inhomogeneous or
anisotropic. The advantage of the EFT framework is that these generalizations
can be systematically incorporated by including additional couplings.
In this section, we explore several possibilities along these lines.

\subsection{Anisotropic materials}

For homogeneous systems, one can still have anisotropy due to a background
field. The retarded response function can then depend on the background.
Given a homogeneous background vector field $\vec{h}_{eff}$, the
Levi-Civita tensor can now be incorporated into the response $\rho^{AB}=\mathscr{D}\epsilon_{\ \ C}^{AB}h_{eff}^{C}$.
Since dissipative effects need to be antisymmetric under the simultaneous
transformation $A\leftrightarrow B$, $\omega\leftrightarrow-\omega$
and $A\leftrightarrow B$ antisymmetry is already included in the
Levi-Civita tensor structure, the response function has to be symmetric
under $\omega\leftrightarrow-\omega$. This means that the leading
order contribution is now independent of $\omega$. \footnote{This is analogous to the dissipative EFT of a spinning black hole
in which case the role of this background is played by the direction
of the spin vector \cite{GoldbergerLiRothstein}.} The corresponding response function is
\begin{equation}
G_{R,diss}^{A,B}(t,\vec{x})=\mathscr{D}\epsilon^{ABC}h_{eff}^{C}\delta(t)\delta^{3}(\vec{x})
\end{equation}
and gives rise to the conservation equation
\begin{equation}
\frac{\partial}{\partial t}\vec{m}=-\lambda\vec{m}\times(\vec{m}\times\vec{h}_{eff}).
\end{equation}
For systems with conserved parity, the external field $\vec{h}_{eff}$
is an effective magnetic field in the sense of being parity odd to
ensure an even-parity response function. 

This damping equation involving an effective external background magnetic
field is known as the Landau-Lifshitz damping equation \cite{LandauLifshitz}.
We observe that from the EFT point of view, the Landau-Lifshitz and
Gilbert damping terms are distinguished by symmetries.

For completeness, we note that in the literature there are generalizations
to Gilbert damping by introducing anisotropic damping tensors. In
field theoretic language, anisotropy corresponds to explicitly breaking
of $SO(3)$ and is thus straightforwardly realized in the EFT by a
spurion condensation. Instead of introducing a symmetry-breaking VEV
to an explicit spurion operator in the action, one may assume the
2-point functions acquire a VEV and the most general resulting dissipative
response function at LO is given by
\begin{equation}
G_{R,diss}^{A,B}(\omega)=S^{AB}\omega+A^{AB},
\end{equation}
where $S^{AB}$ and $A^{AB}$ are general symmetric and antisymmetric
tensors which are not $SO(3)$ invariant. The generalized damping
equation for an anisotropic but homogeneous magnetic system then become
\begin{equation}
\frac{\partial}{\partial t}\vec{m}=-\vec{m}\times\boldsymbol{S}\cdot\frac{\partial}{\partial t}\vec{m}+\vec{m}\times\bm{A}\cdot\vec{m}.
\end{equation}
The exact form of these anisotropy tensors can then be extracted from
the microscopic details of the given full theory.

\subsection{Inhomogeneous materials}

For inhomogenous materials, e.g. configurations with background spin
textures, we may have contributions from higher orders in (spatial)
derivative expansion. The simplest possible operator is given by the
coupling
\begin{equation}
S_{int}=\int d^{4}x\partial_{i}\chi_{a}\tilde{\mathcal{O}}^{ai}(X).
\end{equation}
In terms of the orientation vector, this is equivalent to
\begin{equation}
S_{int}\approx\int d^{4}x\left(\hat{n}\times\partial_{i}\hat{n}\right)\cdot\vec{\mathcal{O}}^{i}(X),\label{eq:deriv}
\end{equation}
at leading order in $\chi$-field.

Again promoting the spin rotation transformation $\delta\hat{n}=\vec{\omega}\times\hat{n}$
to a local parameter $\vec{\omega}(x)$, we find an additional contribution
to the current divergence
\begin{equation}
\delta S_{int}=\int d^{4}x\vec{\omega}\cdot\left(2\partial_{i}\hat{n}\hat{n}+\left(\hat{n}\hat{n}-\delta\right)\cdot\partial_{i}\right)\cdot\vec{\mathcal{O}}^{i},
\end{equation}
where we have used the fact that $\hat{n}\cdot\partial_{i}\hat{n}=\frac{1}{2}\partial_{i}\hat{n}^{2}=0$
to simplify the expression. For the spectral function of the form
\begin{equation}
\rho^{iA,jB}(\omega)=\mathscr{E}\omega\delta^{ij}\delta^{AB},
\end{equation}
this leads to the damping term
\begin{align}
c_2\partial_t\hat{n}= & \mathscr{E}\left(2\partial_{i}\hat{n}\hat{n}\cdot\left(\partial_{t}\hat{n}\times\partial_{i}\hat{n}\right)+\left(\hat{n}\hat{n}-\delta\right)\cdot\partial_{t}\left(\hat{n}\times\vec{\nabla}^{2}\hat{n}\right)\right)\nonumber \\
= & \mathscr{E}\left(2\hat{n}\times\left(\hat{n}\times\partial_{i}\hat{n}\right)\left(\hat{n}\times\partial_{i}\hat{n}\right)\cdot\partial_{t}\hat{n}+\left(\hat{n}\hat{n}-\delta\right)\cdot\partial_{t}\left(\hat{n}\times\vec{\nabla}^{2}\hat{n}\right)\right),
\end{align}
where we have used that $\hat{n}\times\left(\hat{n}\times\partial_{i}\hat{n}\right)=-\partial_{i}\hat{n}$
on the first term. More compactly, this is
\begin{equation}
\frac{\partial}{\partial t}\vec{m}=\vec{m}\times \mathbb{A}\cdot\frac{\partial}{\partial t}\vec{m}+\frac{\mathscr{E}\left(\hat{n}\hat{n}-\delta\right)}{c_2m_s}\cdot\partial_{t}\left(\vec{m}\times\vec{\nabla}^{2}\vec{m}\right),
\end{equation}
where the first term contains the generalized damping tensor
\begin{equation}
\mathbb{A}=\frac{2\mathscr{E}}{c_2 m_s}\left(\vec{m}\times\partial_{i}\vec{m}\right)\left(\vec{m}\times\partial_{i}\vec{m}\right).
\end{equation}
This corresponds to the generalized Gilbert damping in the presence
of non-trivial spin textures (i.e. when $\nabla\vec{m}\neq0$) \cite{ZhangZhang}.

We note that Eq. (\ref{eq:deriv}) is only the leading order derivative
correction to the damping dynamics. The EFT framework is capable of
systematically generating higher derivative corrections. For example,
other types of inhomogeneity may be attributed to interactions in
the lattice model \cite{BrinkerDiasLounis} by
\begin{equation}
\vec{m}_{i}\times\sum_{ij}\mathcal{G}_{ij}\cdot\vec{m}_{j},
\end{equation}
where the $i,j$ indices label the lattice sites associated with the
magnetic moments. In the continuum field theory, the lattice variables
become $\vec{m}_{i}\mapsto\vec{m}(\vec{x}_{i})$ and the tensorial
structure becomes the response function $\mathcal{G}_{ij}\mapsto G_{R,diss}(t,\vec{x}_{i}-\vec{x}_{j})$,
except that unlike the one in Eq. (\ref{eq:resp}), it is non-local
(no longer proportional to $\delta(\vec{x})$). In the simpler case
that the long-range coupling falls off sufficiently quickly, these
terms are traded for a series expansion
\begin{equation}
\vec{m}\times\sum_{n}\mathcal{A}_{i_{1}\dots i_{n}}\partial_{i_{1}\dots i_{n}}^{n}\partial_{t}\vec{m},
\end{equation}
for some coefficient tensors $\mathcal{A}_{i_{1}\dots i_{n}}$. In
terms of the action, this means the spectral functions are now dependent
on the wave vectors
\begin{equation}
\rho^{AB}(\omega,\vec{k})=\sum_{n}\tilde{\mathcal{A}}_{i_{1}\dots i_{n}}k_{i_{1}}\dots k_{i_{n}}\omega\delta^{AB}.\label{eq:inhom}
\end{equation}

\section{Magnon damping term from phonons}

From the magnetoelastic EFT, the generic magnon-phonon couplings are given by \cite{PavaskarPencoRothstein}
\begin{equation}
\mathcal{L}_{ph}=-\frac{1}{2}F_{2}^{ij}(\nabla_{(i}\pi_{j)})\partial_{i}\hat{n}\cdot\partial_{j}\hat{n}+\frac{1}{2}\rho\tilde{F}_{3}(\nabla_{(i}\pi_{j)})D_{t}\hat{n}\cdot D_{t}\hat{n},
\end{equation}
where $D_t\equiv \partial_t+v^i\partial_i$ with the velocity of the material given by $v_i=-\partial_t\phi(D^{-1})^{\ i}_j$. The "full theory" (technically the EFT at the next level of the hierarchy) action constrains the form of the couplings between magnons the external sector to be
\begin{equation}
\mathcal{L}_{ph}=\frac{1}{2}\partial_{i}\hat{n}\cdot\partial_{j}\hat{n}\mathcal{O}_{2}^{ij}(\pi)+\frac{1}{2}\partial_{t}\hat{n}\cdot\partial_{t}\hat{n}\mathcal{O}_{3}(\pi)+\frac{1}{2}\partial_{t}\hat{n}\cdot\partial_{i}\hat{n}\mathcal{O}_{4}^{i}(\pi),
\end{equation}
where the last term arises from the linear-in-$v$ contribution in the  expansion $D_t\hat{n}\cdot D_t\hat{n}$.

For ferromagnets, the dispersion relation dictates the first term to be dominant. After integrating out the external sector using the Schwinger-Keldysh method $\int\mathcal{D}\pi\mathcal{D}\tilde{\pi}$, we find its contribution to the in-in equation of motion is given by
\begin{equation}
c_2\partial_t\hat{n}=\hat{n}\times\partial_{i}\left(\partial_{j}\hat{n}\Bigr\langle\mathcal{O}_{2}^{ij}\Bigr\rangle\right),
\end{equation}
where the in-in expectation value is given by
\begin{equation}
\Bigr\langle\mathcal{O}_{2}^{ij}(x)\Bigr\rangle=\int d^{4}x'\mathcal{G}_{R,2}^{ij,kl}(x-x')\partial_{k}\hat{n}(x')\cdot\partial_{l}\hat{n}(x').
\end{equation}
For 4-index tensor structures under $SO(3)$, there are two invariant tensors corresponding to the symmetric-traceless and trace irreps. Therefore, one may write the leading-order dissipative contribution to the retarded response function as
\begin{equation}
\mathcal{G}_{R,2}^{ij,kl}(x)\simeq\left(\frac{1}{2}\delta^{ij}\delta^{kl}\mathscr{C}_{2}+\delta^{i(k}\delta^{l)j}\mathscr{D}_{2}\right)\delta^{3}(\vec{x})\partial_{t}\delta(t),
\end{equation}
where $\mathscr{C}_2$ and $\mathscr{D}_2$ are the independent (Wilson) coefficients.

Substituting the results, we find the damping equation in a similar form
\begin{align}
&c_{2}\partial_{t}\hat{n}=\frac{1}{2}\mathscr{C}_{2}\hat{n}\times\partial_{i}\left(\partial^{i}\hat{n}\partial_{t}\left(\partial_{j}\hat{n}\cdot\partial^{j}\hat{n}\right)\right)+\mathscr{D}_{2}\hat{n}\times\partial_{i}\left(\partial_{j}\hat{n}\partial_{t}\left(\partial^{i}\hat{n}\cdot\partial^{j}\hat{n}\right)\right)\nonumber\\
=\mathscr{C}_{2}&\hat{n}\times\partial_{i}\left(\partial^{i}\hat{n}\partial^{j}\hat{n}\cdot\partial_{j}\partial_{t}\hat{n}\right)+\mathscr{D}_{2}\hat{n}\times\partial_{i}\left(\partial^{j}\hat{n}\partial^{i}\hat{n}\cdot\partial_{j}\partial_{t}\hat{n}\right)+\mathscr{D}_{2}\hat{n}\times\partial^{j}\left(\partial_{i}\hat{n}\partial^{i}\hat{n}\cdot\partial_{j}\partial_{t}\hat{n}\right),
\end{align}
or more compactly
\begin{equation}
\frac{\partial}{\partial_t}\vec{m}=\vec{m}\times\mathbb{D}\cdot\frac{\partial}{\partial_t}\vec{m},
\end{equation}
where the "damping tensor" $\mathbb{D}$ is given by
\begin{align}
\mathbb{D}=&\frac{1}{c_2 m_s}\Big[\partial_{i}\left(\mathscr{C}_{2}\partial^{i}\hat{n}\partial^{j}\hat{n}+\mathscr{D}_{2}\partial^{j}\hat{n}\partial^{i}\hat{n}\right)+\mathscr{D}_{2}\partial^{j}\left(\partial_{i}\hat{n}\partial^{i}\hat{n}\right)\nonumber\\
&+\left(\mathscr{C}_{2}\partial^{i}\hat{n}\partial^{j}\hat{n}+\mathscr{D}_{2}\partial^{j}\hat{n}\partial^{i}\hat{n}\right)\partial_{i}+\mathscr{D}_{2}\partial_{i}\hat{n}\partial^{i}\hat{n}\partial^{j}\Big]\partial_{j}.
\end{align}

We notice that the form of the couplings restricts the damping tensor
to appear at higher-orders in derivative expansions and hence they
are expected to be small compared with contributions from fermions
(e.g. electrons) in the long wavelength limit. However, for insulating materials that have electron-magnon coupling suppressed, we expect their effects to be more significant.

\section{Conclusion and discussions}

In this paper, we used the in-in (Schwinger-Keldysh) formalism to generalize
the recently constructed EFT of magnetoelasticity \cite{PavaskarPencoRothstein}
to describe damped magnetic dynamics. We discover that the Gilbert
damping term naturally arises as the simplest symmetry consistent
dissipative correction within the in-in formalism. Systematic generalizations
to anisotropic and inhomogeneous setups also yield desired results
such as the Landau-Lifshitz magnetic damping equation. Moreover, we
are able to predict the form of phononic contribution to the damping
dynamics. Thus we have shown that this is a useful framework to derive
dissipative dynamics from first principles and to predict the forms of higher-order corrections in a systematic way.

It would be interesting to investigate the explicit full theory model
of the ``external sector'' such as the fermionic fields in Eq. (\ref{eq:ferm})
and extract the relevant Wilson coefficients by matching the response
functions. In this way, we may gain better insights into what controls
the damping parameter and give more predictive power to the EFT approach.
It would also be interesting to match the relevant coefficients in
Eq. (\ref{eq:inhom}) to obtain an EFT framework for a generalized
class of models. 

Furthermore, various applications of the magnetoelastic
EFT \cite{EspositoPravaskar,PavaskarRothstein} have appeared more recently. It would be interesting to investigate the effects of adding dissipative terms into these problem. There are also further developments in the technical aspects of such EFTs \cite{NicolisRothstein,AkyuzGoonPenco}. It is natural to consider their implications on the non-conservative sector. We leave these problems for future works.

\section*{Acknowledgement}

The author thanks Ira Rothstein for advising throughout
the project and a careful reading of the manuscript. The author also thanks Riccardo Penco for important discussions,
Shashin Pavaskar for other useful discussions, and Witold Skiba for comments on the draft. This work is partially
supported by the grants DE- FG02-04ER41338 and FG02- 06ER41449.

\end{document}